\documentclass[10pt, conference, compsocconf]{IEEEtran}
\usepackage{cite,epsfig,verbatim,rotating,subfigure}
\usepackage{algorithmic}
\usepackage[lined,boxed,commentsnumbered]{algorithm2e}
\usepackage{pgfplots}
\usepackage{tikz}
\usepackage[mediumspace,mediumqspace,Grey,squaren]{SIunits}
\newtheorem{definition}{Definition}

\begin{document}

\title{An Efficient Algorithm for Clustering of Large-Scale Mass Spectrometry Data}

\author{\IEEEauthorblockN{Fahad Saeed$^{1,*}$, Trairak Pisitkun$^1$, Mark A. Knepper$^1$ and Jason D. Hoffert$^1$}
\IEEEauthorblockA{$^1$Epithelial Systems Biology Laboratory\\
National Heart Lung and Blood Institute (NHLBI)\\
National Institutes of Health (NIH)\\
Bethesda, Maryland USA}
}

\maketitle
\begin{abstract}
High-throughput spectrometers are capable of producing data sets containing thousands of spectra for a single biological sample. These data sets contain a substantial amount of redundancy from peptides that may get selected multiple times in a LC-MS/MS experiment. In this paper, we present an efficient algorithm, CAMS (\underline{C}lustering \underline{A}lgorithm for \underline{M}ass \underline{S}pectra) for clustering mass spectrometry data which increases both the sensitivity and confidence of spectral assignment. CAMS utilizes a novel metric, called F-set, that allows accurate identification of the spectra that are similar. A graph theoretic framework is defined that allows the use of F-set metric efficiently for accurate cluster identifications. The accuracy of the algorithm is tested on real HCD and CID data sets with varying amounts of peptides. Our experiments show that the proposed algorithm is able to cluster spectra with very high accuracy in a reasonable amount of time for large spectral data sets. Thus, the algorithm is able to decrease the computational time by compressing the data sets while increasing the throughput of the data by interpreting low S/N spectra.

\end{abstract}

\IEEEpeerreviewmaketitle

\section{Introduction}
\label{sec:intro}

Mass spectrometry based proteomics is an emerging area and has useful applications in biology such as studying the regulation of cellular processes \cite{app1}, cancer molecular therapeutics\cite{app2}\cite{app3} and others\cite{app4}. Mass spectrometry often generates thousand to millions of spectra that needs to be analyzed. The usual computational procedure invoked, after the raw data
is generated from the mass spectrometers is to search the
spectra against a protein database. The algorithms used for searching e.g. Sequest,
Inspect, Xtandem etc, are essentially brute force methods that try to deduce the
peptide from a given spectra. Even algorithms that use
advanced techniques to reduce the computational time e.g. tag-based for Inspect,
two-pass database for X!Tandem etc. are still not computationally efficient enough
for analyzing millions of spectra in a reasonable amount of time.

It is common for the same peptides to get selected for fragmentation multiple times
in a given MS run, making fraction of MS/MS data sets redundant. Searching the same spectra repeatedly, even with computationally efficient tools, wastes a lot of
time and computational resources. The problem is even more pronounced when data from
multiple runs are merged. The redundancy can reach up to $50\%$ for large
data sets \cite{beer,red2,million}. 

The main goal of the work presented in this paper, is to formulate an efficient and accurate
algorithm for clustering of large-scale mass spectrometry data. In order to accomplish the above task, we introduce
a novel metric (called F-set) that can be used for clustering, and a graph theoretic framework that allows us to
use this metric for efficient cluster extraction. The novel algorithm introduced using the graph-theoretic framework has low computational complexity, thus allowing analysis of large datasets.

The rest of the paper is organized as follows. We start with a brief problem statement and
background information relevant to our discussions in section 2. In section 3, we introduce the graph theoretic
framework and the algorithm for efficient extraction of clusters. Section 4 presents the experimental results
and the performance of the algorithm in terms of cluster accuracy, cluster size. Section 5 concludes the paper with discussion and future work.

\section{Problem Statement and Background Information}
Mass spectrometry data is complex and requires sophisticated algorithms to do the data
processing once the raw data from the mass spectrometer is obtained. The raw data
from the mass spectrometer is then fed to various search algorithms e.g. Sequest, Inspect.
These search algorithms do a thorough job of searching the spectra against a known proteome
data base. After the search is complete, each of the spectra is assigned a peptide (or a set of peptides
with different sites of modifications) to which it corresponds.

There are a number of algorithms that have been introduced for clustering mass spectrometry data. Tabb et. al \cite{tabb}, MS2Grouper algorithm\cite{tabb2}, Beer et. al. developed the Pep-Miner algorithm \cite{beer}, Ramakrishnan et. al.\cite{ram}, Dutta et. al. \cite{dutta} and Frank et. al. \cite{million} are to name a few of these algorithms. The objective of this work is to formulate an algorithm that can accurately and efficiently cluster large numbers of spectra, such that the spectra in a given cluster must belong to the same peptide. More formally we define a cluster as follows: 

\begin{definition}
Let there be $N$ number of spectra $S=\{s_1,s_2,\cdots,s_N\}$ and the peptide corresponding to a spectra represented as $P=\{p_1,p_2, \cdots,p_N\}$. Now let the peptide corresponding to a spectra $s_q$ represented by $p_q$ where $q= \{1,\cdots,N\}$.
\end{definition}

\begin{definition}
A distance function $\delta(p_r,p_t)$ where $p_r \in P, p_t \in P$ is defined as the levenstein distance
of the peptides corresponding to the spectra $s_r$ and $s_t$. Now let the number of clusters be $k$ and represented as $K=\{k_1,k_2,\cdots,k_k\}$ such that set $S$ is divided into $k$ subsets. Then, the spectra $s_r$ and $s_t$ where $s_r,s_t \in S$ should belong to the same cluster $k_i$ where $k_i \in K$ , if and only if, $\delta (p_r,p_t)$ =0 where $p_r,p_t \in P$.
\end{definition}

Note, that during clustering of the spectra, the peptides are not known; since the clustering of the spectra
is performed \emph{before} the searching.

\section{Proposed Graph Theoretic Framework and algorithm}

In this section we propose the similarity criteria that we
use for our algorithm and the rationale behind it. We will then introduce
graph theoretic framework that allows us to use the similarity metric in an
efficient way. This is followed by the proposed clustering algorithm.

\subsection{F-set metric}
Although there has been considerable effort in developing algorithms for
spectral data, all of the approaches have been geared towards counting the
number of spectral peaks that are common between two given spectra. This
information is then used to create a similarity index used by the algorithms \cite{beer,million}. It
makes sense to count the number of peaks that are common between two spectra and
use that for similarity indexing. However, noise and other factors such as
compounded spectra can create false positives for similarity. A similarity index that can mitigate these false positives is necessary for an efficient and accurate clustering algorithm.

We introduce F-set metric in this paper for similarity. The basic idea of the metric is as follows: It is possible for a peak to appear at a certain m/z by a random chance. However, it is far less likely for peaks to appear in consecutive succession just by chance. Thus, it makes sense to formulate a similarity metric that counts the \emph{sets} of similar peaks between two given spectra. We formally define the F-set metric below:

\begin{figure}[htb]
\begin{center}
\includegraphics[scale=0.30,angle=0]{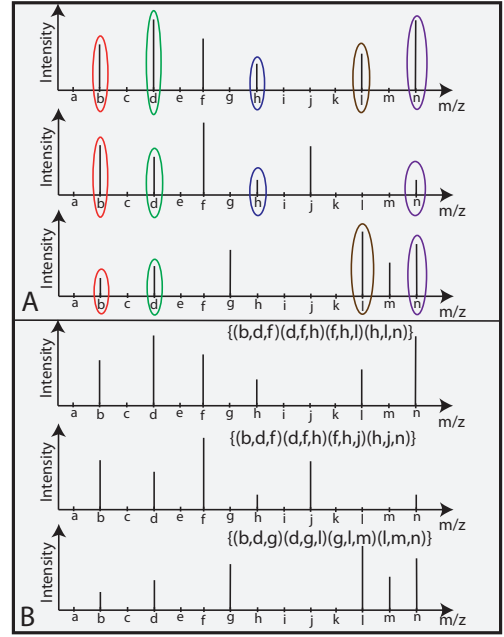}
\caption{\small \label{fig-second} Section A of this figure shows three spectra. The first two spectra map to the same peptides whereas the third spectra maps to a different peptide. Although, the last spectra is not mapped to the same peptide as the first two spectra, we observe significant overlap between the peaks. However, if we make F-sets (of size 3) of the same peaks and spectra, it is clear that the sets formulated do not have much in common for the non-related spectra, and much in common for the spectra that are related, as shown in section B of the figure.}
\end{center}
\end{figure}

\begin{definition}
As before let the spectral data set be represented as $S=\{s_1,s_2,\cdots,s_N\}$. Each spectra has two attributes i.e. m/z and the intensity of the peak. Let there be a fragmentation spectrum $s_j=(m_1,i_1), (m_2,i_2), \cdots,$ $(m_Q,i_Q)$ that is extracted from the mass spectrometry data where $m_d$ represents the m/z ratio and $i_d$ represents the intensity of the peptide at position $d$ and $1 \leq j \leq N$.

Now making sets of peak's at m/z positions of size f. Then creating sets out of the spectra can be presented
as a vector $F(s_i)=\{(m_1 m_2 \cdots m_f),(m_2 m_3 \cdots m_{f+1}), \cdots, (m_{Q-f+1},\cdots,\\m_{Q-1},m_{Q})\}$.
Then the F-set metric calculated for spectra $s_x$ and $s_y$ can be formulated as

\begin{equation}
W(s_x,s_y) = \sum_{i=1}^{|F(s_x)|}\sum_{j=1}^{|F(s_y)|} \phi(F(s_x)[i],F(s_y)[j])
\end{equation}

\begin{equation}
 \phi(a,b) = \left\{ \begin{array}{ll}
         1 & \mbox{if $a[i] = b[j]$}\\
         0 & \mbox{o.w.}\end{array} \right.
\end{equation}

\end{definition}

The F-set, denoted by $W(s_x,s_y)$, can be used as a similarity metric for spectra. The F-set makes set of $m/z$ from the spectra of size $f$ and then compares it with the F-set of the other spectra. If there is a match of a F-set in the other spectra a score of 1 is added. Otherwise a zero score is added. Therefore, the final score $W$ represents the number of F-sets that are common between the two given spectra. The rationale for comparing sets of m/z between two given spectra has to do with the probability of peaks appearing at random places i.e. there is a high probability that a peak would appear at a random place in a spectra due to noise (and hence would result in incorrect clusters if used as a similarity metric), but for peaks to appear in successive order (as sets) for two un-related spectra is less plausible. Figure \ref{fig-second} shows three spectra, of which only two are related. It can be seen from the figure that the F-set metric not only allows distinction between the spectra that are not similar but also allows us to identify spectra that are related i.e. map to the same peptide. Now we formulate the graph theoretic framework to take advantage of F-set metric just defined.

\subsection{Graph Theoretic Framework}
In this section we present the graph-theoretic framework that would allow us to
use F-set metric in an efficient manner.

\begin{definition}
A weighted undirected graph $G=(V,E)$ is a graph where V is a set of vertices and $E \in V \times V$ is a set of edges. Now let a weight $w_{e=(v_i,v_j)} \geq 0$ associated with edge $e=(v_i,v_j)$ where $e \in E$ and $v_i,v_j \in V$.
\end{definition}

 A weighted undirected graph is created with vertices that correspond to each of the spectra. The vertices are connected by weighted edges and each vertex corresponds to a single spectra. The weight on each edge between two given spectra is assigned using the weight calculated using the F-set i.e. the weight assigned to the edge is equal to the F-set calculated between two given spectra. More formally:

\begin{definition}
Given a graph G=(V,E) such that the number of vertices in the graph are equal
to the number of spectra being considered i.e. $|V|=|S|=N$ and an edge connecting
each vertex. Now vertices can be represented by  $V={v_1,v_2, \cdots, v_N}$. Then,
the nodes can be labeled using the following mapping function $\forall v_i \rightarrow s_i$ where
$v_i \in V, s_i \in S, 1\leq i \leq N$. The weight on each edge is the F-set metric that is calculated
for the spectra i.e. $w_e = W(s_i,s_j)$ where $e=(v_i,v_j); s_i,s_j \in S, e\in E, v_i,v_j \in V$.
\end{definition}

\begin{figure}[htb]
\begin{center}
\includegraphics[scale=0.30,angle=0]{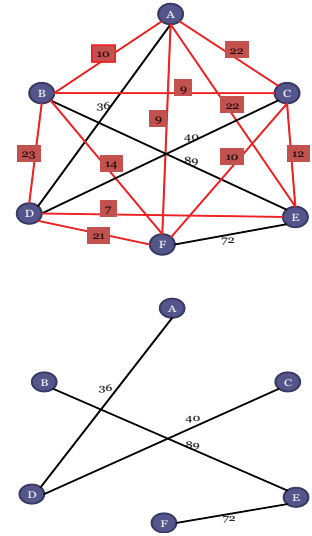}
\caption{\small \label{fig-elimination} The graph from with weighted edges calculated using F-set metric is shown. The value of $\zeta$ is determined using the SVM. Thereafter, the edges having weight less than $\zeta$ are labeled with red boxes (fig a). These edges are then eliminated and the vertices that are still connected are determined using DFS. These connected vertices are reported as potential clusters (fig b).}
\end{center}
\end{figure}

After the above procedure a graph is created that is weighted, and the weight corresponds to the
F-set metric calculated for a given spectra. The next step is to extract the clusters using the graph
that has been created. In order to extract clusters two methods were investigated; one is trivial in which a threshold is chosen by the user; the second threshold is chosen using SVM which our experiments suggested was more effective in chosing the right threshold. After threshold is chosen, the edges that have weight less than threshold are eliminated and the connected components are reported, which can be calculated in $O(V+E)$ time. The algorithm is stated in Algorithm1 and graphical representation of clusters is shown in Fig. \ref{fig-elimination} (b).

\begin{algorithm}
\caption{CAMS}

\begin{algorithmic}
\REQUIRE MS2 spectra data set:
\ENSURE Clusters of spectra such that the cluster has \\spectra that can be mapped to the same peptide:

\begin{enumerate}
\item Read the Sequest search results (.dta) files
\item Enumerate the F-set of a given size for each of the spectra independently
\item For each of the pair determine the F-sets that are common between them
\item Generate the graph using the definition 5 in the \\paper
\item Run SVM on the F-set metrics that gives a $\zeta$ threshold
\item Eliminate the edges that are below the $\zeta$ threshold
\item Determine the vertices that are still connected in \\the graph after elimination
\item Output the vertices that are still connected after elimination as clusters
\end{enumerate}

\end{algorithmic}
\end{algorithm}

\section{Performance Evaluation}
The performance evaluation can be divided into two parts. The first part deals with assessing how good the F-set metric is at distinguishing between related and unrelated spectra. The second part of the evaluation relates to the accuracy of the clusters using the algorithm with different mass spectrometry data sets.

Before we go any further, let us define the quality metric that we use in this paper. The quality
of the clustering can be divided in two parts. The first part is the quality of the individual cluster and the second
is the quality of clustering overall. If we just take an average of the individual quality of the cluster it may
be misleading, since the number of elements in each cluster may be different. Therefore, we defined the
accuracy as a weighted accuracy that allows us to determine the quality of the clustering for each cluster as well
as the overall quality of all clusters. The weighted accuracy is defined as follows:

Assume there are $k$ clusters. Now let the accuracy of a single cluster $i$ be denoted by $a_i$ and the total number of spectra in the cluster be defined as $n_i$ where $1 \leq i \leq k$. Now assume that the number of spectra in a cluster that belong to the same peptide be denoted by $x_i$. Then, the accuracy of a \emph{single} cluster can be defined as :

\begin{equation}
a_i = \frac{x_i}{n_i}
\end{equation}

and the average weighted accuracy (AWA) of the whole dataset under consideration is defined as:

\begin{equation}
AWA = \frac{\sum_{i=1}^{k} a_i n_i}{\sum_{i=1}^{k} n_i}
\end{equation}

AWA takes into account the accuracy of each cluster and gives a global view of the accuracy for a given dataset.

\subsection{Quality assessment}
\subsubsection{Quality with increasing F-set size}
The objective of the first part of quality assessment, is to see how does the quality of the clustering behaves using increasing F-set size. Considering the framework that we introduced in the paper, the increasing size of F-set must correspond to higher accuracy. In order to confirm this, we choose a CID and HCD data sets used in our other studies \cite{phossa}.

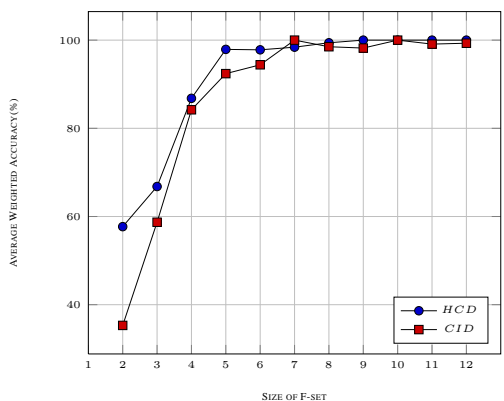
\begin{figure}[!htbp]
\begin{center}

\begin{tikzpicture}[scale=0.8]
    \tikzstyle{every node}=[font=\tiny]
\begin{axis}[grid=major, normalsize,legend pos= south east,legend style={font=\tiny}, xlabel=\textsc{Size of F-set},
ylabel=\textsc{Average Weighted Accuracy(\%)}, xtick={1,2,3,4,5,6,7,8,9,10,11,12}]

\addplot+[black,sharp plot] coordinates
{(2,57.7)(3,66.8)(4,86.8)(5,97.9)(6,97.8)(7,98.4)(8,99.4)(9,100)(10,100)(11,100)(12,100)};

\addplot+[black,sharp plot] coordinates
{(2,35.3)(3,58.7)(4,84.2)(5,92.4)(6,94.4)(7,100)(8,98.5)(9,98.2)(10,100)(11,99.1)(12,99.3)};

\legend{$HCD$,$CID$}

\end{axis}

\end{tikzpicture}
\caption{\small \label{fig-f-set} The average weighted accuracy is shown with increasing F-set size for CID
as well as HCD data sets }
\end{center}
\end{figure}

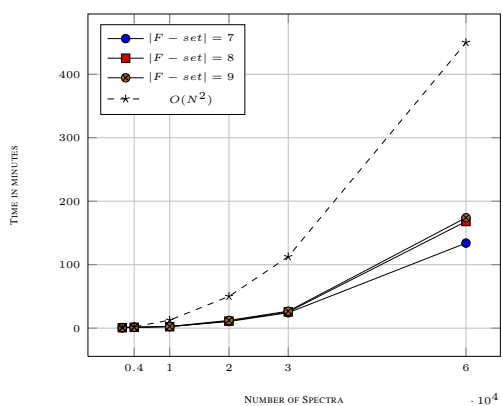
\begin{figure}[!htbp]
\begin{center}

\begin{tikzpicture}[scale=0.8]
    \tikzstyle{every node}=[font=\tiny]
\begin{axis}[grid=major, normalsize,legend pos= north west,legend style={font=\tiny}, xlabel=\textsc{Number of Spectra}, ylabel=\textsc{Time in minutes}, xtick={4000,10000,20000,30000,60000}]

\addplot+[black,sharp plot] coordinates
{(2000,0.5)(4000,1.2)(10000,2.5)(20000,10.5)(30000,24.3)(60000,134)};

\addplot+[black,sharp plot] coordinates
{(2000,0.7)(4000,1.3)(10000,2.3)(20000,11.11)(30000,25.6)(60000,168)};

\addplot+[black,sharp plot] coordinates
{(2000,0.9)(4000,1.5)(10000,2.6)(20000,12.2)(30000,26.7)(60000,174)};

\addplot+[black,dashed] coordinates
{(2000,1)(4000,2)(10000,12.6)(20000,50)(30000,112.5)(60000,450)};

\legend{$|F-set|=7$,$|F-set|=8$,$|F-set|=9$,$O(N^2)$}
\end{axis}
\end{tikzpicture}
\caption{\small \label{fig-timing} The execution time with increasing number of spectra and increasing F-set size are shown. Note that although the CAMS algorithm has a complexity of $O(N^2)$, practically the running times with increasing number of spectra are much less than the theoretical asymptotic times.}
\end{center}
\end{figure}

Fig. \ref{fig-f-set} shows the average weighted accuracy with increasing size of the F-set. In general, the average weighted accuracy increases with increasing F-set size for both CID as well as HCD data sets. The accuracy seems to be leveling off at F-set size of 7 or more. The increase in accuracy can be seen more pronounced in CID data sets as compared to HCD. The HCD data sets have better accuracy with lower F-set size due to better Signal-to-noise ratio as compared to CID. The fact that accuracy increases significantly with increasing F-set size even for CID data sets shows the effectiveness of F-set metric. We see a similar trend with CID and HCD data sets with different conditions as shown in the section below.

\subsubsection{Quality with HCD and CID data sets and complexity analysis}
The data sets that we chose to test the spectral clustering algorithm has been used in other studies\cite{hoffertmcp,phossa}. The data sets consists of CID as well as HCD spectra. The data sets have been produced with varying amount of synthetic AQP-2 peptides. We also use iTRAQ labeled data set from our recent paper \cite{hoffertmcp}. The experiments were conducted with size of F-set equal to $7$. The evaluation of the clustering algorithm with different data sets with varying conditions allows us to assess the performance of the algorithm with "real world" mass spectrometry data sets. Our experiments suggested that the AWA of the clusters obtained  were near $100\%$ accuracy with the minimum accuracy reported as $97.3\%$ (not shown). The time complexity of the algorithm can be shown to be $O(NL^2)+O(c)+ O(N)+ O(V+E) \approx O(N^2)$. As shown in figure \ref{fig-timing}, the execution time with increasing number of spectra is far less than the theoretical $O(N^2)$ execution time and should be expected in practice.

\section{Conclusions and Future Work}
In this paper, we have presented an efficient clustering algorithm suitable for large scale mass spectrometry data. A similarity metric (called F-set) is formulated, and used in the algorithm, based on the spatial locations and intensity of the peaks in a spectra. A graph-theoretic framework is introduced that allows the use of the introduced F-set metric for clustering spectra.  A detailed algorithmic technique based on novel similarity metric (F-set) was described and rigorous time complexity and quality assessment were presented. The graph theoretic framework allows clustering of very large mass spectrometry data sets in a reasonable time. We used CID and HCD data sets with different conditions to assess the quality of the produced clusters. Our experiments suggest that the proposed algorithm allows near-perfect clusters for large-scale mass spectrometry data. The execution time of the algorithm is upper-bounded by $O(N^2)$, but observed execution time is close to linear with increasing number of spectra.

The paper presented is part of the ongoing work on clustering of mass spectrometry data and we plan to expand the work in the future. We would like to investigate both theoretical and application-oriented aspects of clustering large-scale mass spectrometry data. 

\bibliographystyle{abbrv}
\bibliography{mybib}
\end{document}